# Magnetic Brightening of Carbon Nanotube Photoluminescence through Symmetry Breaking

Jonah Shaver[1], Junichiro Kono[1*], Oliver Portugall[2], Vojislav Krstić[2],

Geert L. J. A. Rikken[2], Yuhei Miyauchi[3], Shigeo Maruyama[3], Vasili Perebeinos[4]

*[1]Department of Electrical & Computer Engineering and Carbon Nanotechnology Laboratory, Rice University, Houston, Texas 77005, USA. [2]Laboratoire National des Champs Magnétiques Pulsés, 31400 Toulouse, France. [3]Department of Mechanical Engineering, University of Tokyo, 7-3-1 Hongo, Bunkyo-ku, Tokyo 113-8656, Japan. [4]IBM Research Division, T. J. Watson Research Center, Yorktown Heights, New York 10598, USA.*

**Often a modification of microscopic symmetry in a system can result in a dramatic change in its macroscopic properties. Here we report that symmetry breaking by a tube-threading magnetic field can drastically increase the photoluminescence quantum yield of semiconducting single-walled carbon nanotubes, by as much as a factor of six, at low temperatures. To explain this striking connection between seemingly unrelated properties, we have developed a comprehensive theoretical model based on magnetic-field-dependent one-dimensional exciton band structure and the interplay of strong Coulomb interactions and the Aharonov-Bohm effect. This conclusively explains our data as the first experimental observation of dark excitons 5-10 meV below the bright excitons in single-walled carbon nanotubes. We predict that this quantum yield increase can be made much larger in disorder-free samples.**

Symmetries are behind virtually all physical laws, accompanied by conserved quantities and degenerate quantum states (*1*).    However, truly intriguing physical phenomena sometimes occur when a certain type of symmetry in the system is broken (*2-4*), either spontaneously or by external means.    For example, broken, partial, or incomplete symmetries result in such a diverse range of states/phenomena as Higgs bosons, superconductivity, ferromagnetism, and magneto-resistance due to weak localization.    Here, we show that a symmetry-breaking magnetic field in a single-walled carbon nanotube (SWNT) has enormous influence on a seemingly unrelated macroscopic property of the system: photoluminescence (PL) quantum yield (QY). Specifically, we found that a magnetic field applied parallel to the tube axis drastically increases the PL intensity of semiconducting SWNTs, especially at low temperatures. This phenomenon – *magnetic brightening* – is shown to be a consequence of broken time reversal symmetry, working in tandem with the Aharonov-Bohm effect, and has far-reaching implications for the importance of *symmetry manipulation* as a possible means to modify and control the properties of these unique nano-tubular crystals for optoelectronic device applications.

Carbon nanotubes (*5,6*) possess a variety of novel properties that may find unique applications in nanotechnology.    Semiconducting SWNTs, in particular, have attracted much recent attention for photonic applications (*7*).    They have direct band gaps, which, combined with extreme quantum confinement, are expected to lead to superb optical properties dominated by one-dimensional (1-D) excitons with huge binding energies (*8*).    However, surprisingly, the values of PL QY, i.e., the fraction of absorbed photons re-emitted in fluorescence, reported for SWNT ensembles have been very low ($10^{-3}$ to $10^{-4}$) (*9,10*).    Recent theories (*11-16*) have attributed this low QY to the existence of optically-inactive, or "dark," excitons, *below* the first optically-active, or "bright," exciton state, which can trap much of the exciton population.    The interplay of doubly-degenerate conduction and valence bands coupled with the strong Coulomb interaction characteristic in low-dimensional systems forms this complicated excitonic structure in SWNTs.    Although the excitonic nature of interband optical processes in semiconducting SWNTs has been well established, through, e.g., two-photon PL excitation spectroscopy experiments (*17,18*), conclusive evidence of dark excitons has not yet been reported.

In this paper, we report the first clear evidence of these dark excitons.    We demonstrate, experimentally and theoretically, that a magnetic field can significantly increase the PL QY of semiconducting SWNTs by "brightening" the dark exciton state through temperature and magnetic field dependent PL of pre-aligned, static films of individualized SWNTs.    The PL intensity increased with magnetic field and the amount of brightening increased as temperature was decreased.    We explain the mechanism of *magnetic brightening* by taking into account magnetic-field-dependent effective masses of excitons, population of finite-momentum exciton states, acoustic phonon scattering, and impurity scattering.    A magnetic field applied parallel to the tube axis removes the valley degeneracy by lifting the time reversal symmetry, producing two equally-bright excitonic states at high magnetic fields, whose separation depends on the amount of the Aharonov-Bohm phase, $2\pi\phi/\phi_0$, where $\phi$ is the tube threading magnetic flux and $\phi_0$ is the magnetic flux quantum (*15,19-22*).    This degeneracy lifting mixes different parity excitonic wave functions and provides excitons trapped in the lowest dark state with a radiative recombination pathway, thus brightening the transition and increasing the overall PL QY.

Figure 1 shows a typical data set at 5 K with an excitation wavelength of 780 nm. The pulsed magnetic field and PL spectra are recorded as a function of time, showing intensity increase (magnetic brightening) and peak position red shift [Aharonov-Bohm effect (*20,21*)] tracking the magnetic field magnitude.    Figure 2 shows three sets of PL data taken at different temperatures, with an excitation wavelength of 740 nm, plotted on the same scales.    This figure dramatically demonstrates the difference in intensity increase at different temperatures.    Figure 2a shows 5 K spectra taken at 0 T, 11 T, 23 T, 35 T, 42 T, and 56 T.    At this lowest temperature, the overall intensity increases ~4 to 5 times over the utilized field range.    Figure 2b shows PL spectra at 80 K taken at the same five magnetic fields.    Here we see the amount of brightening to be less than in the 5 K experiment.    Lastly, 260 K spectra are shown in Fig. 2c; the amount of intensity increase is negligible, but there is a slight change in the lineshape indicating the presence of the higher energy peak as previously observed in magneto-absorption data ([20,21]).

Integrated PL intensity versus temperature at various magnetic fields is plotted for the (9,4) nanotube in Fig. 3a.    At 0 T, the integrated PL intensity increases as temperature decreases from 150 K until a peak at ~30 K.    As temperature further decreases from 30 K to 5 K, the intensity decreases.    The non-zero field traces do not show this peak behavior but continue to increase as temperature decreases.    Figure 3b shows integrated PL intensity normalized to the zero field intensity versus magnetic field at different temperatures for the (9,4) nanotube.    This highlights the relative increase in brightening as temperature decreases.    At 200 K, there is little brightening (≤2 times) while the 5 K data shows a large amount of brightening (~5.5 times).

In order to explain the observed magnetic brightening, we have developed a comprehensive theoretical model (see Supporting Online Material for details).    In the presence of time reversal symmetry, i.e., without a magnetic field, exchange-interaction-induced mixing between excitons in two equivalent valleys in momentum space (the K and K' valleys) results in a set of non-degenerate excitonic transitions, only one of which is optically active ([11-16]).    This predicted bright state, however, is not the lowest in energy.    Excitons would be trapped in the dark, lowest-energy state without a radiative recombination path, which can lead to low QY.    When a diameter threading magnetic field is applied to the nanotube long axis, addition of an Aharonov-Bohm phase modifies the circumferential periodic boundary conditions on the electronic wave functions and lifts time reversal symmetry (*15,19-22*).   This symmetry lowering splits the K and K' valley transitions, lessening the exchange-interaction-induced mixing, and causing the recovery of the unmixed direct K and K' exciton transitions, which are both optically active.

Within a simple two-band model for a perfectly aligned nanotube, the relative oscillator strengths for the dark ($\delta$) and bright ($\beta$) states in the presence of a magnetic field may be calculated as

$$I_\delta = \frac{1}{2} - \frac{1}{2}\frac{\Delta_x}{\sqrt{\Delta_x^2 + \Delta_{AB}^2}} \quad \text{and} \quad I_\beta = \frac{1}{2} + \frac{1}{2}\frac{\Delta_x}{\sqrt{\Delta_x^2 + \Delta_{AB}^2}}, \tag{1}$$

respectively.   Here, $\Delta_x$ is the zero-field splitting between the dark and bright exciton states and $\Delta_{AB}$ is the Aharonov-Bohm splitting.   Note that the dark state can be partially brightened by disorder-induced mixing with the bright state; to take into account this contribution, we allow the Aharonov-Bohm splitting to include an additional field-independent value (see Supporting Online Material).   Figure 4a plots the calculated evolution of $I_\delta$ and $I_\beta$ as a function of $\Delta_{AB}$.   As $\Delta_{AB}$ (and thus the magnetic field) increases, the lower state gains oscillator strength at the expense of the higher until both reach equal values at very high magnetic fields where $\Delta_{AB} \gg \Delta_x$. Two bright states of approximately equal oscillator strength appear in high magnetic fields, consistent with previous high magnetic field absorption studies (*21*).

Photoluminescence intensities are not only dependent on oscillator strengths but also on the thermal distribution of exciton population.   In order to accurately model temperature and magnetic field dependent exciton population within and between the exciton bands, we calculated exciton dispersions both in the absence and presence of a magnetic field.    In this model the zero field dispersion for the dark, $i = \delta$, and bright, $i = \beta$, states are given by $E_i(k) = (\Delta_i^2 + \Delta_i \hbar^2 k^2 / m_i)^{1/2}$; where $\Delta_i$ is the energy at the bottom of the band, $m_i$ is the effective mass of the band, $k$ is the wave vector associated with the exciton center-of-mass momentum, and $\hbar$ is the Planck constant divided by $2\pi$. Inclusion of magnetic field results in the modified relations

$$\varepsilon_\delta(k) = \frac{E_\delta(k) + E_\beta(k) - \sqrt{\{E_\delta(k) - E_\beta(k)\}^2 + 4\Delta_{AB}^2}}{2} \qquad (2)$$

$$\varepsilon_\beta(k) = \frac{E_\delta(k) + E_\beta(k) + \sqrt{\{E_\delta(k) - E_\beta(k)\}^2 + 4\Delta_{AB}^2}}{2} \; . \qquad (3)$$

Equations (2) and (3) are plotted at 0 T and 55 T in Fig. 4b.    At 0 T, we calculate that the dark and bright bands have different effective masses [$m_\delta > m_\beta$ (13)] and are separated by an energy of $\Delta_x$ at $k = 0$.    At 55 T, the bands are now separated by energy $(\Delta_x^2 + \Delta_{AB}^2)^{1/2}$ at $k = 0$ and are closer to each other in effective mass.

Light emission is only possible from $k \approx 0$ states in the bright exciton band.    At zero field and room temperature, population of finite-$k$ states in both bright and dark exciton bands exist and restrict the overall amount of PL.    As the sample is cooled, our model predicts that the exciton population distribution would narrow in momentum space, forcing more population to light emitting, $k \approx 0$ states, increasing the PL intensity.    However, as temperature is decreased further, more excitons would populate the optically-inactive, dark exciton band, causing a downturn in PL intensity at the lowest temperatures.    These two competing factors would result in a peak PL intensity at a finite temperature, as observed near 30 K in our 0 T data in Fig. 3a.    Once a magnetic field is applied along the tube axis, excitons trapped in the dark band would gain oscillator strength, allowing for a radiative recombination path and increasing the PL intensity.    This behavior is demonstrated by our non-zero field data in Fig. 3a.

Simulated spectra are shown for 5 K (Fig. 2d), 80 K (Fig. 2e), and 260 K (Fig. 2f) at the same magnetic fields as the corresponding experimental traces in Fig. 2a-2c. We take the magnetic field and temperature dependent exciton bands with relative oscillator strengths and populations, as well as acoustic phonon and impurity scattering, into account to simulate the lineshape (see Supporting Online Material). The 5 K spectra in Fig. 2d successfully reproduce the relative increase in PL intensity as well as the red shift observed as the magnetic field is increased. Since the spectra are dependent on exciton population, and the low temperature restricts the population to the dark state, only one peak is observed. The 80 K spectra in Fig. 2e exhibit the same behavior as 5 K but with less brightening due to a significant population in $k \approx 0$ states of the bright exciton band before the field is applied. The 260 K spectra in Fig. 2f only show weak brightening and an asymmetric lineshape at high fields indicating that at high temperatures excitons are partially populating the higher state. Finally, Fig. 3c shows the calculated PL intensity for the (9,4) nanotube as a function of temperature for different magnetic fields together with the experimental data from Fig. 3a. Our model reproduces the observed trend in the amount of brightening, both in terms of temperature and magnetic field dependence.

Table I shows calculated and fit parameters compiled from fitting low temperature data (5 K) with a series of Voigt (23) peaks. The (10,6), (8,7), (8,6), and (9,4) peaks are the main features in the spectra while the (10,5), (11,3), and (10,2) were fitted as shoulders. Zero-field PL excitation (PLE) maps were examined to determine the approximate peak positions and number of peaks to fit. Preliminary peak positions were determined according to PLE maps and an empirical Kataura plot (24) while the width and shape factor were set to be constant for each field before running the fit. The results of the fitting were combined with chiral index, (n−m) modulo 3 (5,24), family (2n+m) (5), and diameter, as well as the values for the dark-bright splitting deduced, to make Table I.

Our observations and calculations of magnetic brightening clearly demonstrate the existence of previously unobserved dark excitons and their significant influence on the photoluminescence quantum yield of semiconducting SWNTs.    In addition, we show that careful symmetry manipulation by a magnetic field, accompanied by a modification of the circumferential electronic wave functions by the Aharonov-Bohm effect, significantly increases the quantum yield at low temperatures.    Considering the fact that dark excitons are already partially brightened by disorder, we expect the observed signal increase to be even more drastic in disorder-free single-walled carbon nanotube samples.

**Table 1.   Nanotube parameters.**   Chiral indices $n$ and $m$, family ($2n+m$), ($n$-$m$) mod 3, diameter are calculated and indicated based on our tube assignments.    Integrated photoluminescence intensity at 56 T [$I$(56T)] normalized to integrated photoluminescence intensity at 0 T [$I$(0T)] and peak position change $\Delta E$(56T-0T) from 0 T to 56 T were compiled from Voigt fits of the 5 K spectra.    The bright-dark splitting, $\Delta_x$, calculated from fitting the temperature dependence at various fields is also shown.    Bold-faced numbers indicate the four main photoluminescence features while the normal-faced numbers are for shoulder features in the spectra.

| $N$ | $m$ | $2n+m$ | ($n$-$m$) mod 3 | Diameter (nm) | $I$(56T)/ $I$(0T) | $\Delta E$(56T–0T) (meV) | $\Delta_x$ (meV) |
|---|---|---|---|---|---|---|---|
| **10** | **6** | **26** | **1** | **1.11** | **2.13** | **26.9** | **7.7** |
| 10 | 5 | 25 | 2 | 1.05 | 2.95 | 27.9 | |
| **8** | **7** | **23** | **1** | **1.03** | **3.39** | **29.1** | **8.1** |
| 11 | 3 | 25 | 2 | 1.01 | 4.19 | 16.9 | |
| **8** | **6** | **22** | **2** | **0.96** | **6.08** | **27.0** | **7.9** |
| **9** | **4** | **22** | **2** | **0.91** | **5.48** | **24.0** | **5.5** |
| 10 | 2 | 22 | 2 | 0.88 | 7.68 | 22.7 | |

# References and Notes

1. J. J. Sakurai, *Modern Quantum Mechanics* (Addison-Wesley, Redwood City, 1985), Chapter 4.

2. B. Atalay, *Math and the Mona Lisa: The Art and Science of Leonardo da Vinci* (HarperCollins Publishers, New York, 2006).

3. P. W. Anderson, *Concepts in Solids: Lectures on the Theory of Solids* (World Scientific, Singapore, 1997).

4. D. Forster, *Hydrodynamic Fluctuations, Broken Symmetry, and Correlation Functions* (Addison-Wesley, Redwood City, 1975).

5. M. S. Dresselhaus, G. Dresselhaus, Ph. Avouris, eds., *Carbon Nanotubes: Synthesis, Structure, Properties, and Applications,* no. 18 in Topics in Applied Physics (Springer, Berlin, 2001).

6. R. Saito, G. Dresselhaus, M. S. Dresselhaus, *Physical Properties of Carbon Nanotubes* (Imperial College Press, London, 1998).

7. Ph. Avouris, *MRS Bull.* **29**, 403 (2004).

8. T. Ando, *J. Phys. Soc. Jpn.* **66**, 1066 (1997).

9. M. J. O'Connell *et al*., *Science* **297**, 593 (2002).

10. F. Wang, G. Dukovic, L. E. Brus, T. F. Heinz, *Phys. Rev. Lett.* **92**, 177401 (2004).

11. V. Perebeinos, J. Tersoff, Ph. Avouris, *Phys. Rev. Lett.* **92**, 257402 (2004).

12. H. Zhao, S. Mazumdar, *Phys. Rev. Lett.* **93**, 157402 (2004).

13. V. Perebeinos, J. Tersoff, Ph. Avouris, *Nano Lett.* **5**, 2495 (2005).

14. C. D. Spataru, S. Ismail-Beigi, R. B. Capaz, S. G. Louie, *Phys. Rev. Lett.* **95**, 247402 (2005).

15. T. Ando, *J. Phys. Soc. Jpn.* **75**, 024707 (2006).

16. E. Chang, D. Prezzi, A. Ruini, E. Molinari, cond-matt/0603085.

17. F. Wang, G. Dukovic, L. E. Brus, T. F. Heinz, *Science* **308**, 838 (2005).

18. J. Maultzsch *et al*., *Phys. Rev. B* **72**, 241402 (2005).

19. H. Ajiki, T. Ando, *J. Phys. Soc. Jpn.* **62**, 1255 (1993).

20. S. Zaric *et al*., *Science* **304**, 1129 (2004).

21. S. Zaric *et al*., *Phys. Rev. Lett.* **96**, 016406 (2006).

22. J. Kono and S. Roche, "Magnetic Properties," in: *Carbon Nanotubes: Properties and Applications*, edited by M. J. O'Connell (CRC Press, Taylor & Francis Group, Boca Raton, 2006), Chapter 5, pp. 119-151.

23. B. H. Armstrong, *J. Quant. Spectrosc. Radiat. Transfer* **7**, 61 (1967).

24. R. B. Weisman, S. M. Bachilo, *Nano Lett.* **3**, 1235 (2003).

25. This work was supported by the Robert A. Welch Foundation (through Grant No. C-1509), the National Science Foundation (through Grants No. DMR-0134058, DMR-0325474, and No. OISE-0437342), and EuroMagNET under the EU Contract No. RII3-CT-2004-506239 of the 6th Framework "Structuring the European Research Area, Research Infrastructures Action." The authors would also like to thank David J. Hilton for helpful discussions.

*Correspondence and requests for materials should be addressed to J.K. (kono@rice.edu).

# FIGURE CAPTIONS

**Figure 1**   Magnetic brightening in single-walled carbon nanotubes in a pulsed magnetic field. **(a)** Temporal profile of the applied pulsed magnetic field. **(b)** Contour map of near-infrared photoluminescence intensity for a single-walled carbon nanotube film as a function of time and emission wavelength.   Each peak is labelled by its assigned nanotube chirality index ($n$,$m$).   The photoluminescence intensity increases with increasing magnetic field strength. Also note the Aharonov-Bohm-effect-induced red shift of each emission peak with field intensity.   The excitation wavelength was 780 nm and the sample temperature was 5 K.

**Figure 2**   Temperature dependence of magnetic brightening I. Photoluminescence spectra taken with 740 nm excitation at various magnetic fields (0 T, 11 T, 23 T, 35 T, 42 T, and 56 T) at **(a)** 5 K, **(b)** 80 K, and **(c)** 260 K. The amount of brightening is shown to be dramatically higher at lower temperature (scales are equal).   All peaks shift to lower energy with increasing magnetic field at all temperatures, due to the Aharonov-Bohm effect. Simulated photoluminescence spectra for the (9,4) nanotube at **(d)** 5 K, **(e)** 80 K, and **(f)** 260 K at the same fields.   All essential features in the experimental spectra are reproduced: the peak intensity increases (or brightens) with increasing magnetic field, and the amount of brightening is larger at lower temperature.

**Figure 3**   Temperature dependence of magnetic brightening II.   Integrated photoluminescence (PL) intensity of the (9,4) nanotube as a function of **(a)** temperature and **(b)** magnetic field.   The intensity in **(b)**, $I(B)$, is normalized to the zero-field value $I(0)$.   The excitation wavelength was 740 nm.   As temperature is decreased at 0 T, the PL intensity peaks near ~30 K.   The non-zero field traces do not show this peak behavior but continue to increase as temperature decreases.   At high temperatures, there is little brightening (≤2 times) while data at low temperatures show a large amount of brightening (~5.5 times).   **(c)** Temperature dependence of the photoluminescence intensity for the (9,4) peak: comparison between experimental data at three magnetic fields (0 T, 23 T, and 56 T) and theory.

**Figure 4**   Magneto-exciton bands and magnetic field dependence of relative oscillator strengths. **(a)** Calculated relative oscillator strength of the dark ($I_\delta$) and bright ($I_\beta$) excitons vs. Aharonov-Bohm splitting, $\Delta_{AB}$, for a 1 nm diameter pristine nanotube in a parallel field.   Both $I_\delta$ and $I_\beta$ approach ~0.5 at $\Delta_{AB}$ greater than ~50 meV.   **(b)** Calculated dispersions for bright and dark exciton bands at 0 and 55 T for a perfectly aligned nanotube.   At 0 T the higher energy, lower mass ($0.02m_e$), bright exciton is separated from the lower energy, higher mass ($0.2m_e$), dark exciton by 5.5 meV.   At 55 T both peaks are bright and approach similar masses, separated by 55 meV.

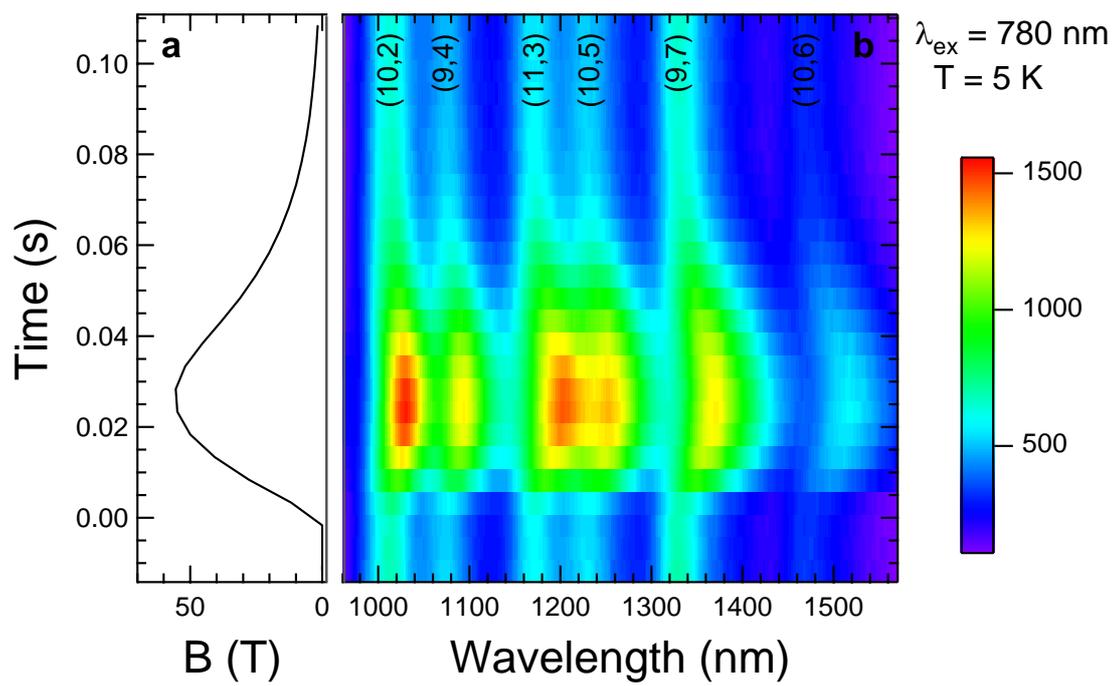

Fig. 1

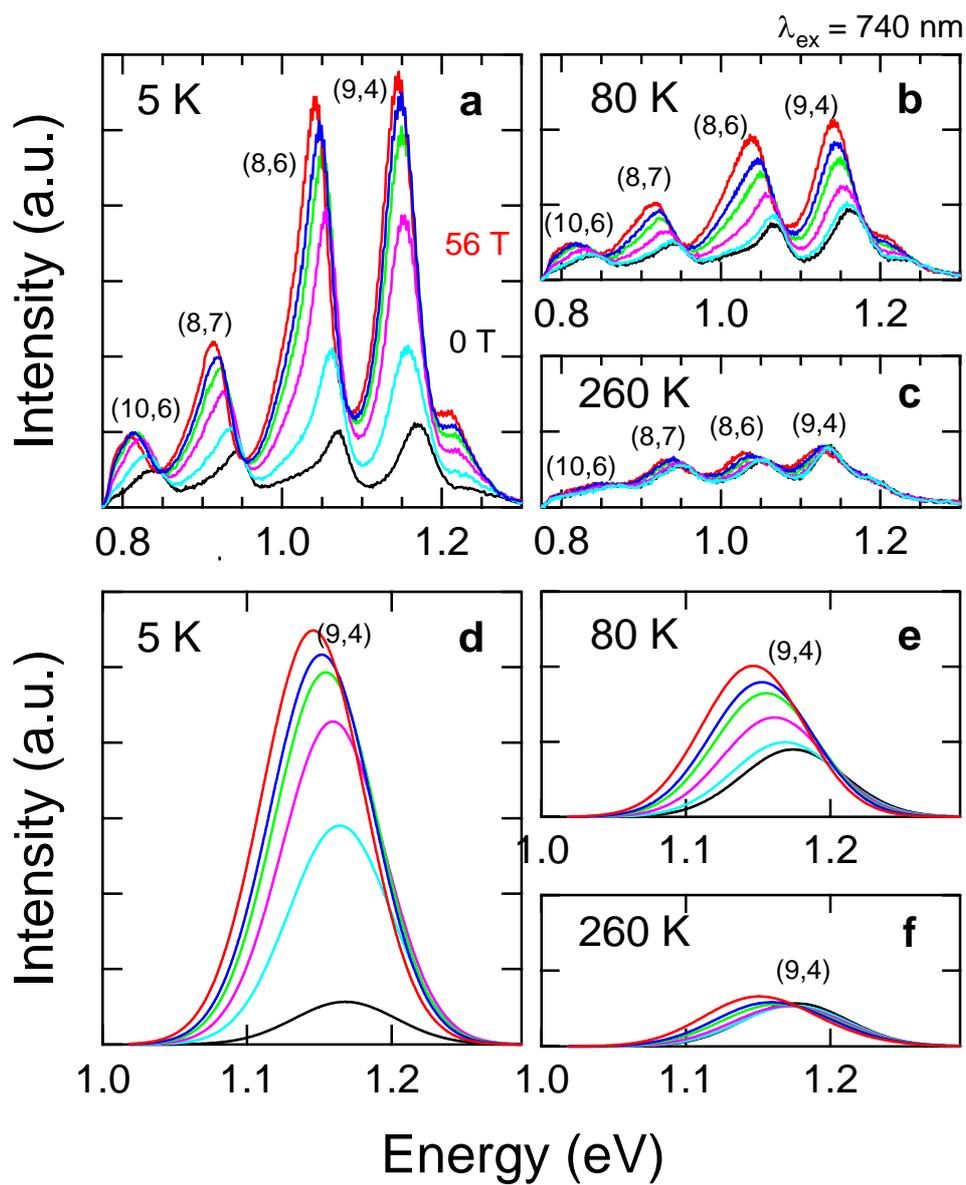

$\lambda_{ex} = 740$ nm

**a** 5 K (10,6) (8,7) (8,6) (9,4) 56 T 0 T

**b** 80 K (10,6) (8,7) (8,6) (9,4)

**c** 260 K (10,6) (8,7) (8,6) (9,4)

**d** 5 K (9,4)

**e** 80 K (9,4)

**f** 260 K (9,4)

Intensity (a.u.)

Energy (eV)

Fig. 2

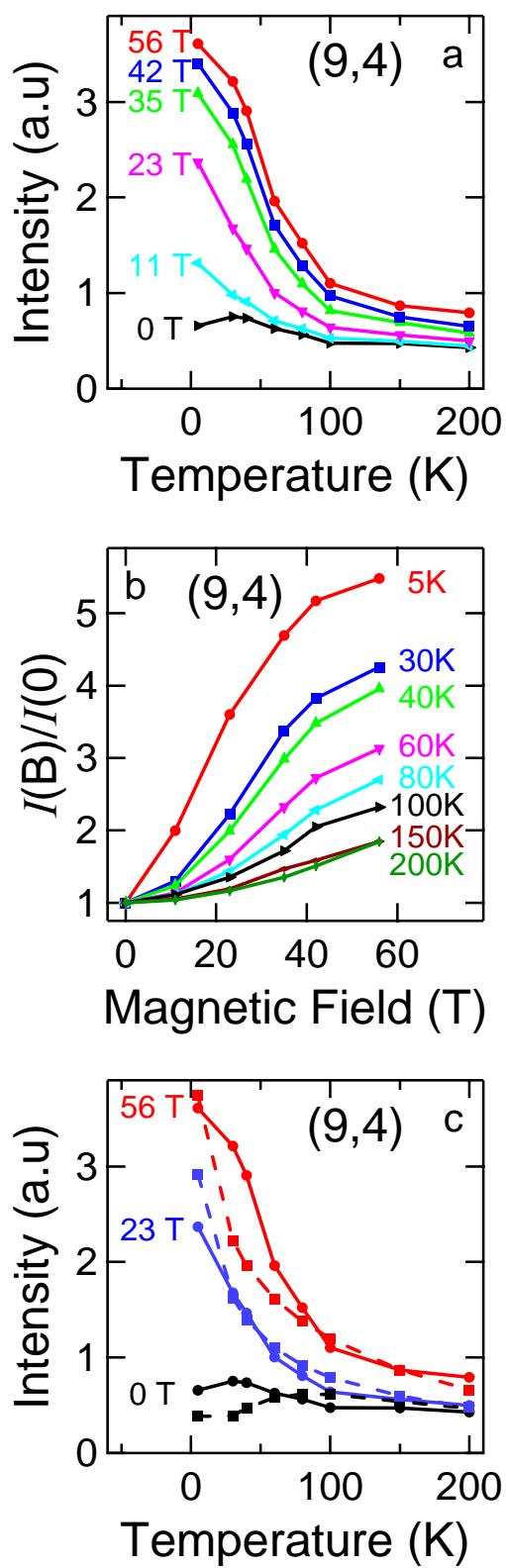

Fig. 3

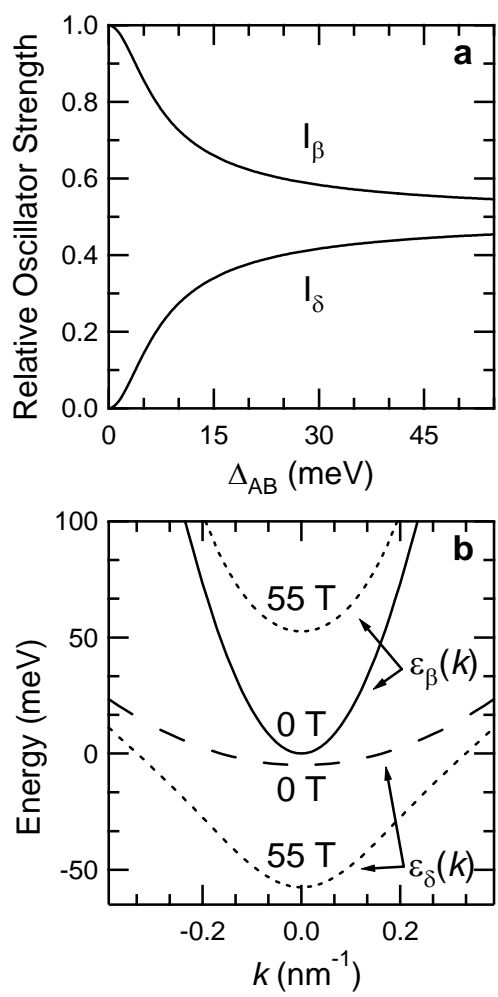

Fig. 4

# Supporting Online Material

## Experimental Methods

In previous magneto-optical studies of carbon nanotubes in aqueous suspensions (*S1-S3*), it was necessary to account for the magnetic alignment dynamics of SWNTs in data analysis.   Although magnetic alignment was advantageous for achieving higher amounts of magnetic flux threading the tubes at high magnetic fields, in this study we are closely following the PL intensity as a function of magnetic field and any alignment effects would greatly complicate data analysis.   In order to avoid this issue, we have made samples that are both static and aligned.   Samples measured in this study were High Pressure CO (HiPco) process SWNTs (batch HPR104).   The SWNTs were first suspended in sodium dodecyl benzene sulfonate (SDBS) using standard techniques before dispersion in bovine protein gelatin (*S4,S5*).   The gelatin was then cast onto an adhesive polyvinyl chloride substrate, stretched, and stuck to a flat surface while the gelatin set and water evaporated.   Upon drying, the film was removed and soaked in alcohol before stretching again.   Linear dichroism measurements show the minimum order parameter (*S6*) to be ~0.18.   Samples show high PL signal and well-defined absorption peaks.

Temperature-dependent magneto-photoluminescence experiments were performed from 5 K to 260 K in pulsed magnetic fields up to 56 T at the Laboratoire National des Champs Magnétiques Pulsés in Toulouse, France.   The magnetic field was generated by a 100 ms, 3 MJ pulse with a 25 ms sinusoidal rise time and an exponential fall-off.   Sample films were measured in a fiber-coupled, reflection, Voigt geometry probe with the nanotube alignment direction parallel to and the light propagation vector perpendicular to the magnetic field.   Spectra were taken with a Roper OMA-V InGaAs diode array every ~5 or 15 ms with a 1 ms collection time in order to maximize field resolution.    The Ti:Sapphire excitation laser was turned off in between data collection periods in order to minimize heating.

**Theory of Magnetic Brightening**

The doubly degenerate valence and conduction bands in single-walled carbon nanotubes give rise to four singlet electron-hole pair excitations, or excitons.    The lowest-energy exciton has zero circumferential angular momentum but even spatial symmetry, and thus, it cannot decay radiatively (i.e., "dark").    The odd symmetry, zero angular momentum exciton state has a higher energy due to the Coulomb exchange interaction and it can decay radiatively (i.e., "bright").    The remaining two, higher energy, exciton states have a finite circumferential angular momentum, and thus cannot decay radiatively.

An applied magnetic field breaks time reversal symmetry and mixes the wave functions of the dark and bright excitons of zero circumferential angular momentum. The wave function mixing redistributes the oscillator spectral weight between the two excitons, which leads to the brightening of the lowest energy dark exciton.    In addition, nanotube impurities can break the symmetry (*S7*), which would also result in the wave function mixing and brightening of the dark exciton.    To conveniently represent both the magnetic-field-induced mixing and disorder-induced mixing, we assume that the 'Aharonov-Bohm' splitting ($\Delta_{AB}$) consists of a term proportional to the magnetic field flux ($\phi = \pi B d^2/4$) and a term proportional to the strength of disorder ($\Delta_{dis}$): $\Delta_{AB} = \mu\phi + \Delta_{dis}$, where $B$ is the magnetic field strength, $d$ is the nanotube diameter, and $\mu$ is a proportionality constant for the Aharonov-Bohm splitting induced by the magnetic field. Note that $\mu$ depends on the angle between the tube axis and the magnetic field direction, and in the case of an ensemble of nanotubes with some angular distribution, $\mu$ depends on the degree of alignment (i.e., the nematic order parameter, *S*).

In the absence of symmetry breaking interactions ($\Delta_{AB} = 0$), the energy dispersions of the dark ($i = \delta$) and bright ($i = \beta$) excitons are given by $E_i(k) = (\Delta_i^2 + \Delta_i \hbar^2 k^2/m_i)^{1/2}$, where $\Delta_i$ is the energy at the bottom of the band, $m_i$ is the effective mass of the band, and $k$ is the wave vector associated with the exciton center-of-mass momentum. In the presence of mixing ($\Delta_{AB} \neq 0$), the eigenvalues of the Hamiltonian for each $k$-point are

$$\varepsilon_\delta(k) = \frac{E_\delta(k) + E_\beta(k) - \sqrt{\{E_\delta(k) - E_\beta(k)\}^2 + 4\Delta_{AB}^2}}{2} \qquad (1a)$$

$$\varepsilon_\beta(k) = \frac{E_\delta(k) + E_\beta(k) + \sqrt{\{E_\delta(k) - E_\beta(k)\}^2 + 4\Delta_{AB}^2}}{2}. \qquad (1b)$$

The spectral weights of the dark and bright excitons are

$$I_\delta = \frac{1}{2} - \frac{1}{2}\frac{\Delta_x}{\sqrt{\Delta_x^2 + \Delta_{AB}^2}}; \quad I_\beta = \frac{1}{2} + \frac{1}{2}\frac{\Delta_x}{\sqrt{\Delta_x^2 + \Delta_{AB}^2}}. \qquad (2)$$

where $\Delta_x = \Delta_\beta - \Delta_\delta$ is the dark-bright energy splitting in zero magnetic field. In the absence of a magnetic field, the effective mass of the bright exciton is predicted to be much smaller that that of the dark exciton (*S7*). It is interesting to note that the magnetic field dependence of the effective mass $m_i^*(B)$ follows from Eqs. (1a) and (1b) as

$$\frac{1}{m_\delta^*(B)} = \frac{I_\beta}{m_\delta^*(B=0)} + \frac{I_\delta}{m_\beta^*(B=0)}; \quad \frac{1}{m_\beta^*(B)} = \frac{I_\delta}{m_\delta^*(B=0)} + \frac{I_\beta}{m_\beta^*(B=0)}. \qquad (3)$$

Thus, at very high magnetic fields both their masses and oscillator strengths become equal, as schematically shown in Fig. 4.

The absorption spectrum consists of two peaks, i.e.,

$$\sigma(E) = \frac{I_\delta}{\pi}\frac{\gamma_\delta(E)}{[E - \varepsilon_\delta(0)]^2 + \gamma_\delta^2(E)} + \frac{I_\beta}{\pi}\frac{\gamma_\beta(E)}{[E - \varepsilon_\beta(0)]^2 + \gamma_\beta^2(E)}, \qquad (4)$$

where $\gamma_i$ is the imaginary part of the self-energy of the dark and bright excitons, respectively, which determine the lineshape of the spectra (*S8*). We include two interactions in evaluating the exciton self-energy: (i) acoustic phonon scattering and (ii) impurity scattering.   Using a self-consistent Born approximation, the imaginary part of the self-energy is found to be

$$\gamma_i(E) = \frac{a_0^2 \sqrt{3}}{4\pi d}\left[ g_{ph}k_{\mathrm{B}}T\int dk \frac{-1}{\pi}\mathrm{Im}\{G_i(E-\hbar\omega_k,k)\} + g_{imp}^2\int dk \frac{-1}{\pi}\mathrm{Im}\{G_i(E,k)\}\right] \qquad (5)$$

where $i = \beta, \delta$ and

$$G_i(E,k) = \frac{1}{E - \varepsilon_i(k) + i\gamma_i(E)} \qquad (6)$$

is the exciton Green function, $g_{ph}$ and $g_{imp}$ are the exciton-acoustic phonon and exciton-impurity coupling strengths, respectively, $T$ is temperature, $k_{\mathrm{B}}$ is the Boltzmann constant, $\omega_k = v_s k$ is the acoustic phonon frequency, where $v_s = 2.1 \times 10^6$ cm/s is the sound velocity, and $a_0$ is the graphene lattice constant.   We neglect the renormalization of exciton energy due to interaction with impurities and phonons (*S9*).

To calculate emission spectra, we assume a thermal distribution of excitons. The emission spectral lineshape is also given by Eq. (4), except that $\gamma_i$ in the numerators are evaluated with integrals as in Eq. (5), but weighted with the statistical factors, $\exp(-\varepsilon_i(k)/k_{\mathrm{B}}T)/Z(T)$.   Note that $\gamma_i$ in the denominators of Eq. (4) are not modified. The statistical sum $Z(T)$ has three contributions from all four lowest energy exciton bands, i.e.,

$$Z(T) = \frac{a_0^2 \sqrt{3}}{4\pi d}\left[\int dk \exp\left(-\frac{\varepsilon_\delta(k)}{k_{\mathrm{B}}T}\right) + \int dk \exp\left(-\frac{\varepsilon_\beta(k)}{k_{\mathrm{B}}T}\right) + 2\int dk \exp\left(-\frac{E_\alpha(k)}{k_{\mathrm{B}}T}\right)\right], \qquad (7)$$

where $E_\alpha(k)$ is the energy dispersion of the dark exciton above the bright state.   The energy of finite angular momentum (intervalley) electron-hole pairs has been found only very weakly dependent on magnetic field (*S10*), so we model the dark state above the bright exciton state by a magnetic field independent effective mass approximation: $E_\alpha(k) = E_\beta(k) + \Delta_\alpha$.

The best fit to the integrated intensity for the (9,4) tube in Fig. 6 was obtained with the following set of parameters: $\Delta_\delta = 1.169$ eV, $\Delta_x = \Delta_\beta - \Delta_\delta = 5.5$ meV, $m_\delta = 0.2 m_e$, $m_\beta = 0.02 m_e$, $\Delta_{dis} = 2.1$ meV, $g_{imp} = 0.1$ eV, $\mu = 0.64$ meV/Tesla-nm$^2$, $g_{ph} = 10$ eV, and $\Delta_\alpha = 20$ meV.    The quantitative agreement between the theory and experiment is good.    With the same set of parameters we were also able to reproduce in Fig. 2d-f the lineshape of the emission spectra shown in Fig. 2a-c.    The spectra in Fig. 2d-f are Gaussian-broadened with a width of 30 meV to account for the inhomogeneous broadening of the ensemble of tubes.

*References for supporting online material:*

S1.    S. Zaric *et al.*, *Science* **304**, 1129 (2004).

S2.    S. Zaric *et al.*, *Nano Lett.* **4**, 2219 (2004).

S3    S. Zaric *et al.*, *Phys. Rev. Lett.* **96**, 016406 (2006).

S4.    Y. Miyauchi, Y. Murakami, S. Chiashi, S. Maruyama, paper presented at the 27$^{th}$ Fullerene-Nanotubes General Symposium, 29 (2004).

S5.    S. Kazaoui *et al.*, *Appl. Phys. Lett.* **87**, 211914 (2005).

S6.    J. R. Lakowicz, *Principles of Fluorescence Spectroscopy* (Plenum Publishing Corp., New York, ed. 3, 2006).

S7.    V. Perebeinos, J. Tersoff, Ph. Avouris, *Nano Lett.* **5**, 2495 (2005).    In zig-zag tubes, the symmetry operation of the mirror plane reflection perpendicular to the tube axis gives rise to the even and odd spatial parity sets of states.    In chiral tubes, the symmetry operation of the interchange of the two atoms in the primitive unit cell leads to the similar classification of the eigenstates.    The two symmetry operations coincide in zig-zag tubes.

S8.    Y. Toyozawa, in *Excitonic Processes in Solids*, edited by M. Ueta, H. Kanzaki,
       K. Kobayashi, Y. Toyozawa, and E. Hanamura (Springer, Berlin, 1986).

S9.    V. Perebeinos, J. Tersoff, Ph. Avouris, *Phys. Rev. Lett*. **94,** 027402 (2005).

S10.   T. Ando, *J. Phys. Soc. Jpn.* **75**, 024707 (2006).